\documentclass[conference]{IEEEtran}
\IEEEoverridecommandlockouts
\usepackage{cite}
\usepackage{amsmath,amssymb,amsfonts}
\usepackage{algorithmic}
\usepackage{graphicx}
\usepackage{textcomp}
\usepackage{xcolor}
\usepackage{hyperref}
\usepackage{bm}
\usepackage{subcaption}
\usepackage{multirow}
\usepackage{fancyhdr}

\def\BibTeX{{\rm B\kern-.05em{\sc i\kern-.025em b}\kern-.08em
    T\kern-.1667em\lower.7ex\hbox{E}\kern-.125emX}}
\begin{document}

\title{Attention-Based SINR Estimation in User-Centric Non-Terrestrial Networks\\
}

\author{\IEEEauthorblockN{Bruno De Filippo\IEEEauthorrefmark{1}, Alessandro Guidotti\IEEEauthorrefmark{1}\IEEEauthorrefmark{2}, Alessandro Vanelli-Coralli\IEEEauthorrefmark{1}}
\IEEEauthorblockA{\IEEEauthorrefmark{1}Department of Electrical, Electronic, and Information Engineering (DEI), Univ. of Bologna, Bologna, Italy}
\IEEEauthorblockA{\IEEEauthorrefmark{2}National Inter-University Consortium for Telecommunications (CNIT), Bologna, Italy}
\{bruno.defilippo, a.guidotti, alessandro.vanelli\}@unibo.it}

\maketitle
\thispagestyle{fancy}

\begin{abstract}
The signal-to-interference-plus-noise ratio (SINR) is central to performance optimization in user-centric beamforming for satellite-based non-terrestrial networks (NTNs). Its assessment either requires the transmission of dedicated pilots or relies on computing the beamforming matrix through minimum mean squared error (MMSE)-based formulations beforehand, a process that introduces significant computational overhead. In this paper, we propose a low-complexity SINR estimation framework that leverages multi-head self-attention (MHSA) to extract inter-user interference features directly from either channel state information or user location reports. The proposed dual MHSA (DMHSA) models evaluate the SINR of a scheduled user group without requiring explicit MMSE calculations. The architecture achieves a computational complexity reduction by a factor of three in the CSI-based setting and by two orders of magnitude in the location-based configuration, the latter benefiting from the lower dimensionality of user reports. We show that both DMHSA models maintain high estimation accuracy, with the root mean squared error typically below 1 dB with priority-queuing-based scheduled users. These results enable the integration of DMHSA-based estimators into scheduling procedures, allowing the evaluation of multiple candidate user groups and the selection of those offering the highest average SINR and capacity.\end{abstract}

\begin{IEEEkeywords}
Non-Terrestrial Networks, User-Centric Beamforming, Self-Attention, Deep Learning
\end{IEEEkeywords}

\section{Introduction}\label{ch:1_intro}
The signal-to-interference-plus-noise ratio (SINR) of a communication link plays a major role in the effectiveness of the data exchange. On the one hand, the achievable rate in communication systems is limited by the SINR through the Shannon formula, effectively limiting the amount of bits that can be sent per time and bandwidth unit \cite{bib:distanceMadocJournal}. On the other hand, procedures in communication systems rely on SINR-related metrics to take decisions, \textit{e.g.}, handovers from one cell to another can be carried out once the measured reference signal received quality overcome a certain threshold \cite{bib:handoverNTN}. Such proxy for link quality can typically be measured at the terminal using reference signals; however, especially in the context of satellite-based non-terrestrial networks (NTNs), complications can be introduced during the reporting phase: geostationary-class satellites are characterized by a large propagation delay which can result in outdated reports, \textit{e.g.}, in the case of user mobility, while non-geostationary-class satellites at lower orbit claim lower latency but introduce aging effects due to their inherent orbital movement.
Focusing on user-centric beamforming NTN systems, the assessment of the SINR strongly relies on the set of users scheduled during the same time slot. Indeed, once a group of users is scheduled and the beamforming matrix is computed, either by means of minimum mean squared error (MMSE) or location-based MMSE (LB-MMSE), the SINR achieved by each user can be evaluated (considering the theoretical clear sky channel formula based on the user’s location in case of LB-MMSE). However, such evaluation is based on the MMSE equation, which introduces a great computational overhead \cite{bib:CFMIMO}. Thanks to the function approximation capabilities of neural networks, the SINR achieved by a group of scheduled users in a user-centric beamforming NTN system can be assessed with reduced computational complexity by 1) extracting complex features from the users' channel vectors, and 2) evaluating the interactions between different users' features. In this context, we identified the popular attention mechanism as a potential candidate for extracting inter-user interference-related information. To the best of our knowledge, previous works have either obtained formulations of the SINR in cell-free MIMO systems based on the MMSE equation \cite{bib:CFMIMO}\cite{bib:SINRup} or focused on the distribution of the SINR \cite{bib:SINRdist}; on the opposite, approaches to SINR estimation based on deep learning (DL) are underexplored, especially in the NTN framework. We believe that the availability of low-complexity SINR estimates can improve user scheduling algorithms by providing an effective evaluation metric to select the best user group to be scheduled among a set of candidate groups. Based on these considerations, in this paper we:
\begin{itemize}
	\item Propose a low-complexity self-attention-based SINR estimator for user-centric beamforming;
	\item Assess the performance of the model in a NTN considering random and priority-based scheduling.   
\end{itemize}

\section{System model}
\begin{figure}
	\centering
	\includegraphics[width=0.83\columnwidth]{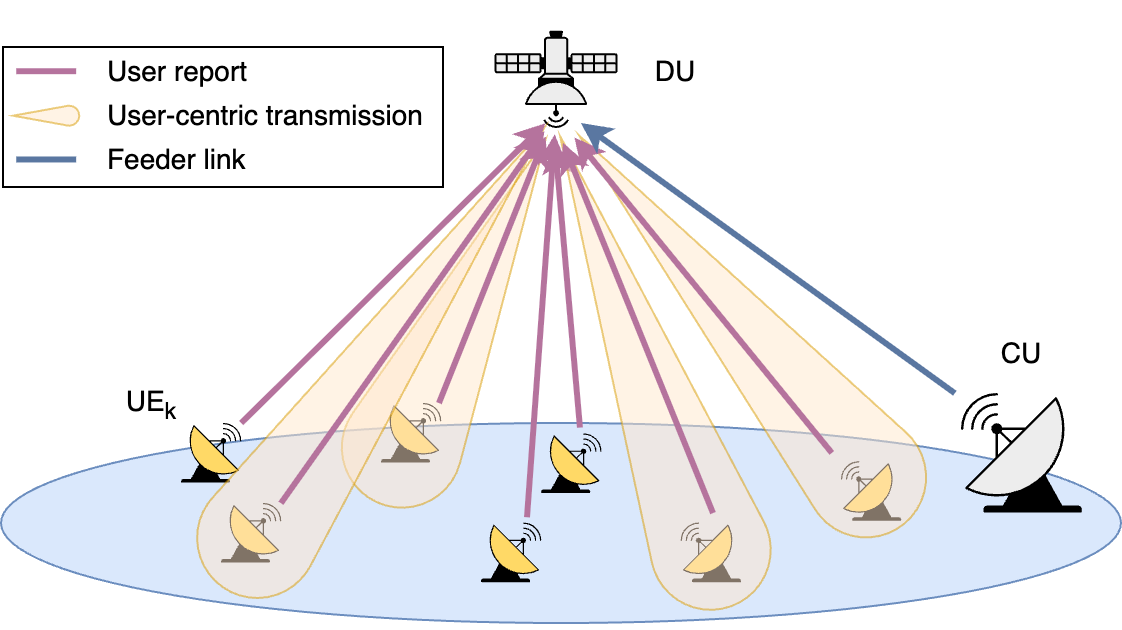}
	\caption{Considered system model.}
	\label{fig:SINRest_systemModel}
\end{figure}
We here consider a low Earth orbit (LEO) satellite providing coverage to a set of $N_{UE}$ users in the service area in a user-centric fashion (Figure \ref{fig:SINRest_systemModel}). The spaceborne node is equipped with a regenerative payload, with the on board gNodeB (gNB) distributed unit (DU) implementing up to the upper medium access control functions on board (\textit{i.e.}, functional split option 4), including user scheduling; furthermore, the satellite is equipped with a direct radiating array composed of $N_R$ radiating elements capable of generating up to $N_B$ simultaneously active beams \cite{bib:STARDUST_D3.2}. At least one gateway, implementing the gNB centralized unit (CU), is always reachable by the satellite. When in visibility, each user, equipped with a very-small-aperture terminal, periodically reports either its estimated CSI or location (latitude and longitude) to the gNB. The channel coefficient between the $k$-th user equipment (UE) and the $n$-th radiating element at time $t$ can be computed as \cite{bib:CFMIMO}:
\begin{equation}\label{eqn:channel}
    \mathbf{H}_{k,n}^{(t)} = \frac{g_{k,n}^{(tx, t)}g_{k,n}^{(rx, t)}}{\frac{4\pi d_k^{(t)}}{\lambda}\sqrt{L_k^{(t)}\kappa B_k T_k}}e^{-j\frac{2\pi}{\lambda}d_k^{(t)}},
\end{equation}
where $g_{k,n}^{(tx, t)}$ and $g_{k,n}^{(rx, t)}$ represent the transmitting and receiving gain, respectively, between the $k$-th UE and the $n$-th radiating element; $d_k^{(t)}$ is the $k$-th user's slant range; $L_k^{(t)}$ represents all stochastic losses (\textit{e.g.}, due to shadowing and the atmosphere); $\kappa$ is Boltzmann's constant; $B_k$ and $T_k$ represent the $k$-th user's allocated bandwidth and equivalent noise temperature, respectively; and $\lambda$ is the wavelength corresponding to the carrier frequency $f_c$. When new reports are available at the gNB, the user's estimated channel vectors (either reported in case of CSI reports or computed by means of \eqref{eqn:channel} with $L_k^{(t)}=0$ dB in case of location reports) are grouped together in the channel matrix $\hat{\mathbf{H}}_{UE}^{(t)} \in \mathbb{C}^{N_{UE}\times N_R}$; at every scheduling instant, occurring with periodicity $T_{sched}$, such channel matrix is then used to select one group of users for each time slot of duration $T_{sched}^{(slot)}$ in the scheduling window. Thus, in the generic time slot $t$, a subset of $N_{UE, sched}^{(t)} \leq N_B$ users are scheduled for transmission, resulting in the estimated scheduled channel matrix  $\hat{\mathbf{H}}^{(t)} \in \mathbb{C}^{N_{UE, sched}^{(t)}\times N_R}$. The corresponding beamforming matrix can be computed through the MMSE equation \cite{bib:CFMIMO}:
 \begin{equation}\label{eqn:mmse}
 	\mathbf{B}^{(t)} = (\hat{\mathbf{H}}^{(t)})^H \left( \hat{\mathbf{H}}^{(t)} (\hat{\mathbf{H}}^{(t)})^H + \alpha \mathbf{I}_{N_B} \right)^{-1},
 \end{equation}
where $\alpha=N_R/P_{av}$ is the beamforming regularization factor, with $P_{av}$ representing the total transmission power at the spaceborne node, and $\mathbf{I}_{a}$ representing the $a \times a$ identity matrix. The per-antenna constraint normalization ensures that all antenna elements transmit the same amount of power:
 \begin{equation}
 	\tilde{\mathbf{B}}^{(t)} = \sqrt{\frac{P_{av}}{N_R}}diag\left( \frac{1}{\lvert \lvert \mathbf{B}_{1,:}^{(t)}\rvert \rvert}, \dots, \frac{1}{\lvert \lvert \mathbf{B}_{N_R,:}^{(t)}\rvert \rvert}\right)\mathbf{B}^{(t)},
 \end{equation}
with $diag(x_1,\dots,x_a)$ being the $a \times a$ diagonal matrix having values $x_1, \dots, x_a$ on the main diagonal. The SINR of user $k$ at time slot $t$ can then be evaluated as:
\begin{equation}\label{eqn:SINR}
	SINR_k^{(t)}=\frac{SNR_k^{(t)}}{1+INR_k^{(t)}}=\frac{\lvert\lvert \mathbf{H}_{k,:}^{(t)}\tilde{\mathbf{B}}_{:,k}^{(t)} \rvert\rvert^2}{1+\sum_{l=1, l\neq k}^{N_{UE, sched}^{(t)}} \lvert\lvert \mathbf{H}_{k,:}^{(t)}\tilde{\mathbf{B}}_{:,l}^{(t)} \rvert\rvert^2}.
\end{equation}

\section{Dual multi-headed self-attention for SINR estimation}
\subsection{DMHSA Model}\label{ch:SINRest_DMHSA}
\begin{figure*}
    \centering
    \includegraphics[width=0.95\textwidth]{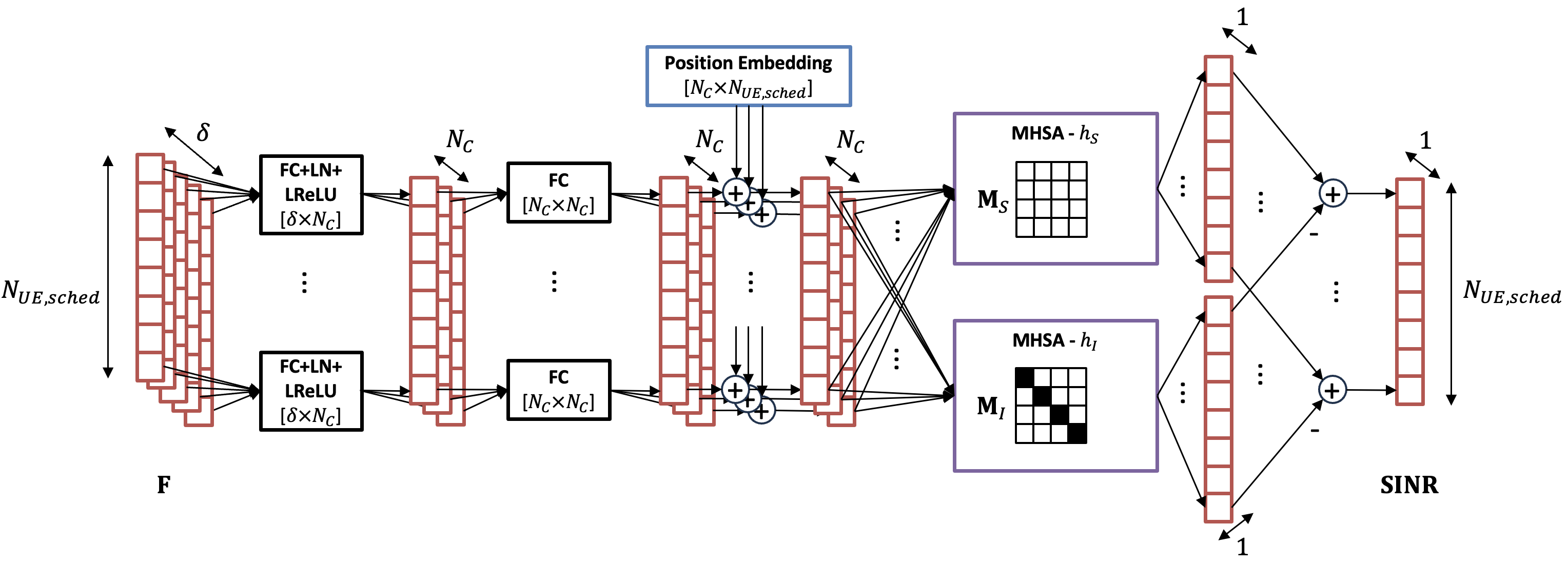}
    \caption{DMHSA SINR estimation model.}
    \label{fig:SINRest_DMHSA}
\end{figure*}
The proposed DL model for SINR estimation is a dual multi-headed self-attention (DMHSA) architecture (Figure \ref{fig:SINRest_DMHSA}). For the sake of conciseness, we recall only the core concepts relevant to the model. Scaled dot-product attention (SDPA) maps queries, keys, and values to output representations by assigning weights proportional to the dot-product between queries and keys. In self-attention, these vectors are linearly projected from the same input, and the multi-headed formulation (MHSA) applies several parallel projections to capture different relationships within the data. In the proposed model, the attention mask allows selective control over which user interactions contribute to each attention head.

\paragraph*{Input representation}
Dropping the temporal index for clarity, the model's input is a matrix $\mathbf{F}\in\mathbb{R}^{N_{UE,sched}\times \delta}$ containing $\delta$ features per scheduled user. Since $N_{UE,sched}$ varies across samples, a binary mask $\mathbf{m}^{(DMHSA)}=[\mathbf{1}_{1\times N_{UE,sched}},\mathbf{0}_{1\times (N_B-N_{UE,sched})}]\in \{0,1\}^{1\times N_B}$ is provided to ensure efficient parallelization. The network then computes $\hat{\mathbf{Y}}^{(DMHSA)} = f^{(DMHSA)}(\mathbf{F},\mathbf{m}^{(DMHSA)})$, where each row $\mathbf{F}_{k,:}\in\mathbb{R}^{\delta}$ corresponds to the features of user $k$. For CSI-based beamforming, such features include:
\begin{enumerate}
    \item The normalized phases of the $k$-th estimated CSI vector, $\bm{\Phi}_{k,n} = \frac{\angle \hat{\mathbf{H}}_{k,n}}{\sigma_\varphi} \forall n=1,\dots,N_R$, with $\sigma_\varphi=\frac{\pi}{\sqrt{3}}$, under the assumption $\angle\hat{\mathbf{H}}_{k,n}\sim U(-\pi, \pi)$;
    \item The normalized mean squared magnitude of the CSI, $\bm{\psi}_k= \left(  \frac{1}{N_R} \sum_{n=1}^{N_R} \left| \hat{\mathbf{H}}_{k,n} \right|^{2} - \mu_H \right) / \sigma_H$, with $\mu_H$ and $\sigma_H$ being standardization parameters empirically estimated from a data sample; and
    \item The ratio $\rho = N_{UE,sched}/N_B$.
\end{enumerate}
The resulting feature vector has size $\delta^{(CSI)}=N_R + 2$:
\begin{equation}
\mathbf{F}_{k,:}^{(\mathrm{CSI})} =
    \left[
        \bm{\Phi}_{k,1}
        \ldots,
        \bm{\Phi}_{k,N_R},
        \bm{\psi}_k,
        \rho
    \right],
\end{equation}

For location-based beamforming, the input features are the user coordinates in the $(u,v)$ reference system, $u_k$ and $v_k$, and the same ratio $\rho$, yielding $\delta^{(GEO)}=3$:
\begin{equation}
    \mathbf{F}_{k,:}^{(\mathrm{GEO})} = 
    \left[
        u_k,
        v_k,
        \rho
    \right].
\end{equation}
It is worth mentioning that the inclusion of the ratio $\rho =N_{UE,sched}/N_B$ provides the network with coarse information about the amount of interference to be expected.

\paragraph*{Embedding extraction}
Each user’s feature vector is processed independently by a stack of fully connected (FC) layers with $N_C$ neurons, layer normalization (LN), and leaky rectified linear unit activation. Position embedding (PE) is then added to add further information on the expected level of interference.

\paragraph*{Dual attention mechanism}
The resulting embeddings are fed to two parallel MHSA+FC modules. Both operate on the set of scheduled users, but they differ in how user interactions are masked:
\begin{itemize}
    \item The SNR module uses a full attention mask $\mathbf{M}_S=\mathbf{1}_{N_{UE,sched}\times N_{UE,sched}}$ to enable all interactions.
    \item The INR module uses $\mathbf{M}_I=\mathbf{1}_{N_{UE,sched}\times N_{UE,sched}}-\mathbf{I}_{N_{UE,sched}}$, excluding self-interference to focus solely on inter-user contributions.
\end{itemize}
Each module uses $h_S=h_I=h=4$ heads, with projection dimensions chosen such that $d_k=d_v=N_C/h=2$. The standardized SINR in dB is obtained by subtracting the output of the second module's FC layer with single output neuron from the first's: intuitively, they correspond to an estimation of the terms $(1+INR_k)$ and $SNR_k$ in dB in \eqref{eqn:SINR}.

\subsection{Computational complexity}
Regardless of the choice between CSI-based beamforming and location-based beamforming, the SINR can be evaluated through \eqref{eqn:SINR} (with the theoretical CSI being computed based on the users’ geographical location in case of location-based beamforming). When assessing the MMSE beamforming algorithm’s complexity from \eqref{eqn:mmse}, which is at the base of the SINR evaluation, two main sources of complexity can be identified:
\begin{itemize}
    \item The matrix multiplication between the CSI matrix $\hat{\mathbf{H}}^{(t)}\in\mathbb{C}^{N_{UE,sched}\times N_R}$ and its conjugate transpose with complexity $\mathcal{O}\left(N_{UE,sched}^2 N_R\right)$; and
	\item The inverse operation on the $N_{UE,sched}\times N_{UE,sched}$ complex matrix $\hat{\mathbf{H}}^{(t)}\left(\hat{\mathbf{H}}^{(t)}\right)^H+ \alpha\mathbf{I}_{N_{UE,sched}}$, with complexity $\mathcal{O}\left(N_{UE,sched}^3 \right)$.
\end{itemize}
However, as the number of scheduled users $N_{UE,sched}$ is typically limited with respect to the number of radiating elements $N_R$, \textit{i.e.}, $N_{UE,sched}\ll N_R$, the MMSE complexity reduces to:
\begin{equation}
    \mathcal{O}_{MMSE}=\mathcal{O}\left(N_{UE,sched}^2 N_R\right).
\end{equation}

The computational complexity of the DMHSA model can be mainly attributed to FC and MHSA layers:
\begin{itemize}
	\item The first FC layer embeds $\delta$ features into a vector of size $N_C$ for each user, resulting in a computational complexity $\mathcal{O}\left(\delta N_{UE,sched} N_C\right)$.
	\item The second FC layers further refines the $N_C$ features for each user, resulting in a computational complexity $\mathcal{O}\left(N_{UE,sched} N_C^2\right)$.
	\item The MHSA layers apply linear projections to compute the queries, keys, and values matrices; the subsequent computations in the SDPA involve a matrix multiplication between queries and keys; finally, the heads outputs are concatenated to form a vector of $N_C=h\cdot d_k$ features and passed through one final FC layer. Thus, the overall computational complexity of the MHSA layers is \cite{bib:attentionIsAllYouNeed}:
    \begin{equation}
        \mathcal{O}_{MHSA}=\mathcal{O}\left(N_{UE,sched}^2 N_C+N_{UE,sched} N_C^2 \right). 
    \end{equation}
	\item The last pair of FC layers generates one feature from a vector of length $N_C$ for each user, resulting in a computational complexity $\mathcal{O}\left(N_{UE,sched} N_C\right)$.
\end{itemize}
Thus, the computational complexity of the DMHSA model is:
\begin{multline}
    \mathcal{O}_{DMHSA}=\mathcal{O}\left(\delta N_{UE,sched} N_C \right)+\\
    \mathcal{O}\left(N_{UE,sched}^2 N_C+N_{UE,sched} N_C^2 \right).
\end{multline}
In case of CSI-based beamforming, $\delta^{(CSI)}=N_R+2$; thus, the computational complexity becomes $\mathcal{O}\left(N_{UE,sched} N_C N_R \right)+\mathcal{O}\left(N_{UE,sched}^2 N_C+N_{UE,sched} N_C^2 \right)$. Considering that $N_{UE,sched}\ll N_R$, the complexity reduces to $\mathcal{O}\left(N_{UE,sched}N_C N_R+N_{UE,sched} N_C^2 \right)$. Furthermore, we choose $N_C=8$ so that $N_C \ll N_R$; thus, the overall computational complexity for SINR estimation with the CSI-DMHSA model is:
\begin{equation}
    \mathcal{O}_{DMHSA}^{(CSI)}=\mathcal{O}\left(N_{UE,sched} N_C N_R \right),
\end{equation}
which, given $N_C<N_{UE,sched}$ (typically $N_{UE,sched}=N_B=24$), implies that the complexity is effectively reduced with respect to the evaluation through the computation of the MMSE beamforming equation.
In case of location-based beamforming, $\delta^{(GEO)}=3$; thus, the computational complexity becomes $\mathcal{O}\left(N_{UE,sched}^2 N_C+N_{UE,sched} N_C^2 \right)$. With the same consideration on $N_C$ as above, the complexity of the GEO-DMHSA model reduces to:
\begin{equation}
    \mathcal{O}_{DMHSA}^{(GEO)}=\mathcal{O}\left(N_{UE,sched}^2 N_C\right),
\end{equation}
providing a great computational complexity reduction for SINR estimation. The results of the computational complexity assessment are summarized in Table \ref{tab:SINRest_complexity}.
\renewcommand{\arraystretch}{1.6}
\begin{table}[t]
    \caption{Computational complexity for SINR estimation.}
    \label{tab:SINRest_complexity}
    \centering
    \begin{tabular}{|l|c|}
        \hline
        \textbf{SINR estimation method} & \textbf{Computational complexity} \\
        \hline
        MMSE-based                      & $\mathcal{O}_{MMSE}=\mathcal{O}\left(N_{UE,sched}^2 N_R\right)$ \\
        \hline
        CSI-DMHSA                       & $\mathcal{O}_{DMHSA}^{(CSI)}=\mathcal{O}\left(N_{UE,sched} N_C N_R \right)$ \\
        \hline
        GEO-DMHSA                       & $\mathcal{O}_{DMHSA}^{(GEO)}=\mathcal{O}\left(N_{UE,sched}^2 N_C\right)$ \\
        \hline
    \end{tabular}
\end{table}
\begin{figure}
    \centering
    \includegraphics[width=0.7\columnwidth]{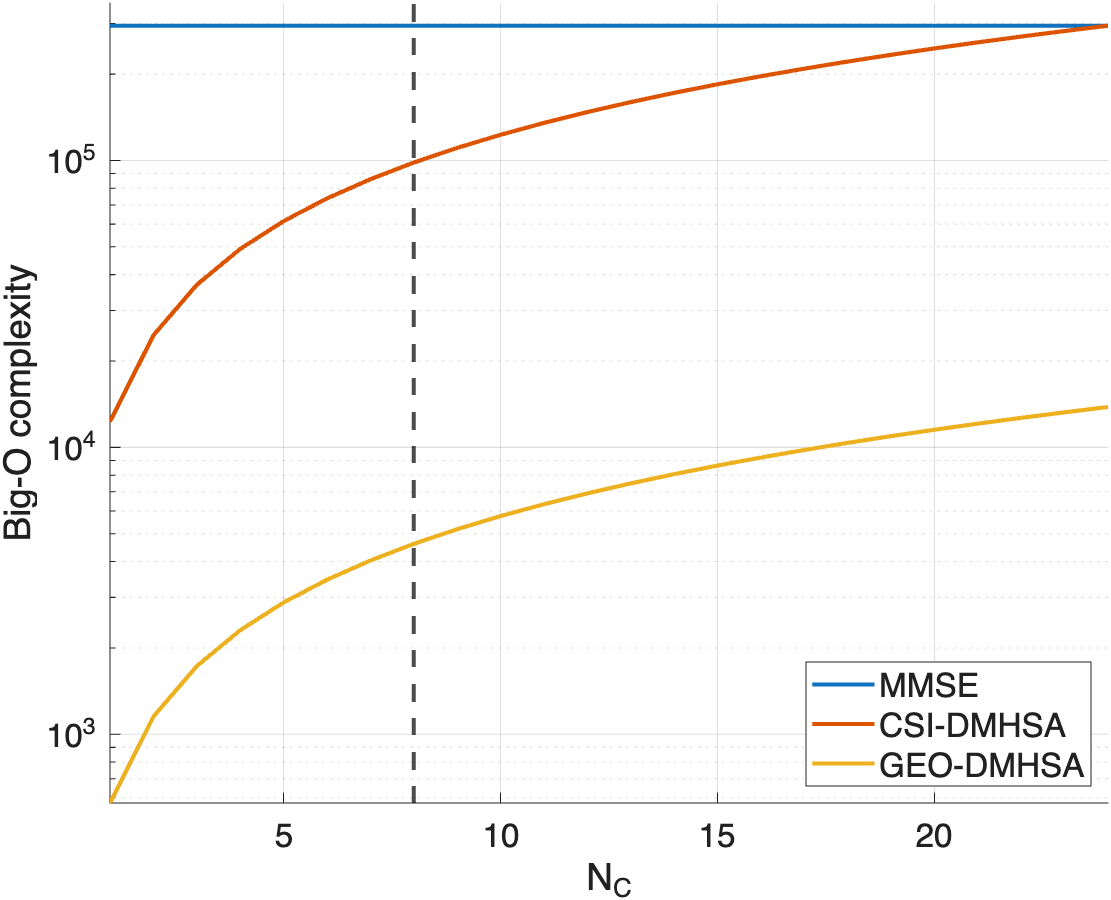}
    \caption{Computational complexity as a function of $N_C$ with $N_{UE, sched}=24$ (black dashed line corresponds to $N_C=8$).}
    \label{fig:SINRest_complexity}
\end{figure}

Figure \ref{fig:SINRest_complexity} reports the Big-O computational complexity achieved by the two DMHSA models (CSI-based and location-based in orange and yellow, respectively) and the MMSE baseline (blue line) as a function of the number of channels $N_C$, \textit{i.e.}, the size of the embeddings of each user’s features vector; the black dashed line represents the value set in the trained models, \textit{i.e.}, $N_C=8$. The plot shows that the reduced features vector in the GEO-DMHSA model ensures that the complexity remains orders of magnitude lower than that of MMSE-based SINR estimation: at $N_C=8$, the location-based algorithm achieves a complexity of $\approx4.6\cdot10^3$ against the $\approx2.9\cdot10^5$ required by the benchmark. The CSI-DMHSA model also ensures a reduction in complexity to $\approx9.8\cdot 10^4$; however, the large features vector makes so the complexity in the CSI-based case scales much faster than in the location-based case with respect to $N_C$: while longer embeddings could improve the SINR estimation accuracy, the increased complexity may disfavour the application of the CSI-DMHSA method. This is not the case with GEO-DMHSA, where even with embeddings of 24 elements the complexity is at one order of magnitude under the MMSE-based SINR estimation. It should also be mentioned that the CSI-DMHSA and GEO-DMHSA models require 4.8k and 762 learnable parameters, respectively, making them extremely compact.

\subsection{Model training}
Two separate models with the reported DMHSA architecture have been trained using CSI reports as input data for one of them and location reports for the other. The models have been trained by simulating scheduling instances at random samples of a satellite pass, using the parameters reported in Table \ref{tab:SINRest_parameters}. At each instance $i$, $N_{UE,sched,\text{ }i}\sim U(8,N_B)$ users are randomly selected for scheduling, their CSI vectors (or location vectors) are pre-processed and collected as training input data sample $\mathbf{F}$ (Section \ref{ch:SINRest_DMHSA}), and their SINR is assessed by means of the MMSE equation and stored as label (\textit{i.e.}, true SINR value) for the corresponding input data sample. Once the batch is filled with $N_{batch}$ samples, the input data $\mathbf{X}^{(DMHSA)}\in\mathbb{R}^{N_B\times \delta \times N_{batch}}$ and the corresponding padding masks $\mathbf{M}^{(DMHSA)}\in\mathbb{R}^{N_B\times N_{batch}}$ are fed to the model, and the corresponding SINR estimates $\hat{\mathbf{Y}}^{(DMHSA)}\in\mathbb{R}^{N_B\times N_{batch}}$ are obtained; then, the MSE loss function $\mathcal{L}_{MSE}$ is evaluated between the SINR estimates at the output of the model and the corresponding ground truth labels $\mathbf{SINR}\in\mathbb{R}^{N_B\times N_{batch}}$:
\begin{gather}
    \mathbf{MSE}_{u,i}= \left( \hat{\mathbf{Y}}^{(DMHSA)}_{u,i} - \mathbf{SINR}_{u,i} \right)^2,\\
    \mathcal{L}_{MSE} = \frac{\sum_{i=1}^{N_{batch}} \sum_{u=1}^{N_B} \mathbf{M}^{(DMHSA)}_{u,i} \odot \mathbf{MSE}_{u,i}}{\sum_{i=1}^{N_{batch}} \sum_{u=1}^{N_B} \mathbf{M}^{(DMHSA)}_{u,i}},
\end{gather}
where $\odot$ is the element-wise product.

\renewcommand{\arraystretch}{1.2}
\begin{table}[t]
    \centering
    \begin{tabular}{|l|c|}
        \hline
        \textbf{Parameter} & \textbf{Value} \\
        \hline
        Subcarrier spacing & $\Delta f=120$ kHz \\
        \hline
        Carrier frequency & $f_c=20$ GHz \\
        \hline
        User bandwidth & $B=190.08$ MHz  \\
        \hline
        Scenario & Rural \cite{bib:STARDUST_D4.5} \\
        \hline
        Channel model & 3GPP NTN system-level \cite{bib:tr38.811}\cite{bib:tr38.821} \\
        \hline
        Number of radiating elements & $N_{R}=512$ \\
        \hline
        Power per radiating element & $P_{av,el}=65$ mW \\
        \hline
        Minimum user elevation angle & 30° \\
        \hline
        NTN node altitude & 1000 km \\
        \hline
        Training batch size & $N_{batch}=8192$ \\
        \hline
        Maximum training epochs & 15000 epochs \\
        \hline
        L2 regularization factor & $\epsilon_R=10^{-6}$ \\
        \hline
        LR warm-up period & $T_w=40$ epochs \\
        \hline
        LR cosine annealing period & $T_c=100$ epochs \\
        \hline
        Minimum LR & $\lambda_{min}=10^{-4}$ \\
        \hline
        Maximum LR & $\lambda_{min}=5\cdot 10^{-3}$ \\
        \hline
        Early stopping patience & $T_s=4$ cosine annealing cycles \\
        \hline
        \end{tabular}
    \caption{Simulation parameters for DMHSA training.}
    \label{tab:SINRest_parameters}
\end{table}

Through the backpropagation algorithm and the Adam optimizer \cite{bib:adam}, the trainable weights of the model are updated, and a new batch of data starts being generated. The training process employs L2 regularization, early stopping, learning rate (LR) warmup, and LR cosine annealing with warm restarts, with Table \ref{tab:SINRest_parameters} reporting the parameters employed to train both of the DMHSA models. A new batch of data is generated at each epoch by means of simulation, and samples contained in previous batches are not reused in the following epochs: for this reason, a validation set is not generated, with early stopping being set based on the training set loss.

\section{Results}
\subsection{Random scheduler}
The objective of the first evaluation is to determine the SINR estimation error under general conditions: \textit{e.g.}, training on randomly scheduled users ensures that the models observe both cases of optimal scheduling and ill-conditioned beamforming matrices. For this reason, the models were first evaluated on a test set under the same conditions of the training set, \textit{i.e.}, each sample contains a random number of randomly scheduled users’ reports obtained in random instants of the satellite pass, for a total of $\approx4\cdot 10^7$ SINR estimates. As for the training data, the simulation parameters used to generate test data are the same reported in Table \ref{tab:SINRest_parameters}. Notably, user locations are sampled from a non-uniform distribution based on the Joint Research Center Global Human Settlement Layer data \cite{bib:GHSL} according to the preprocessing procedure detailed in \cite{bib:STARDUST_D4.5}.

\begin{figure}
     \centering
     \begin{subfigure}[b]{0.48\columnwidth}
         \centering
         \includegraphics[width=\columnwidth]{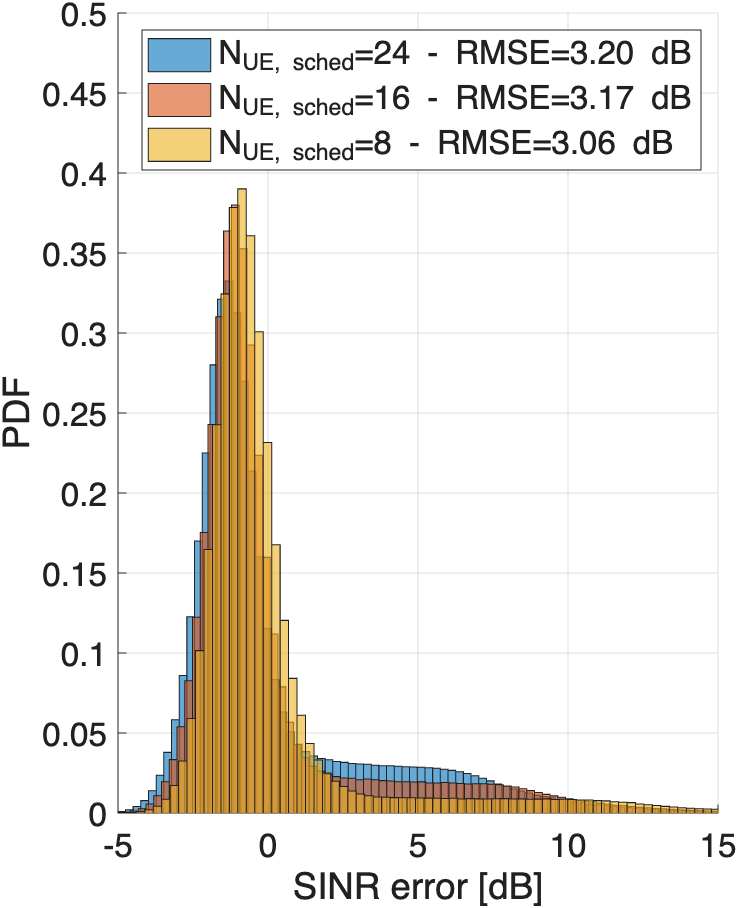}
         \caption{CSI-DMHSA}
         \label{fig:SINRest_errorDistUsers_CSI}
     \end{subfigure}
     \hfill
     \begin{subfigure}[b]{0.48\columnwidth}
         \centering
         \includegraphics[width=\columnwidth]{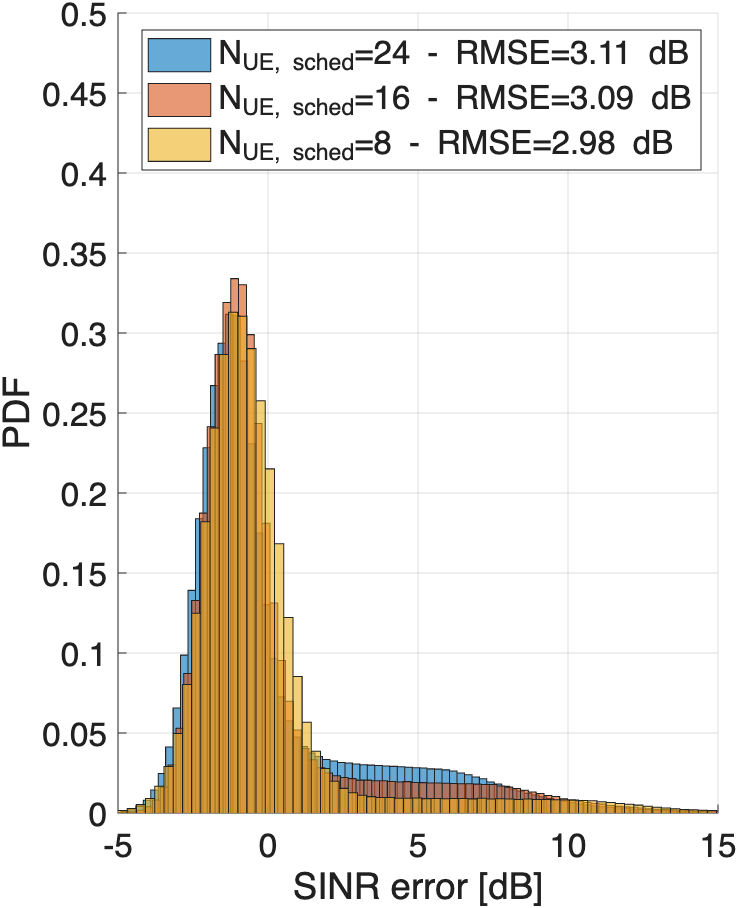}
         \caption{GEO-DMHSA}
         \label{fig:SINRest_errorDistUsers_GEO}
     \end{subfigure}
     \hfill
    \caption{SINR estimation error distribution for different $N_{UE,sched}$ values (random scheduler).}
    \label{fig:SINRest_errorDistUsers}
\end{figure}

Figure \ref{fig:SINRest_errorDistUsers} reports the histogram of the error distribution for both CSI-DMHSA and GEO-DMHSA, computed as $\mathbf{E}=\hat{\mathbf{Y}}^{(DMHSA)}-\mathbf{SINR}$, with probability density function (PDF) normalization. The assessment is carried out by separating the test datasets based on the number of scheduled users. The distribution is skewed in both cases, suggesting that the models cannot provide accurate estimates in certain conditions: this is the case of ill-conditioned beamforming matrices resulting from poor user scheduling (\textit{i.e.}, the random scheduler often selects users that are geographically close together, resulting in highly correlated CSI vectors). As expected, for both models, the root MSE (RMSE) increases with the number of users scheduled, and so does the frequency of high-error estimations. It should be noted that the high SINR error tails are caused by poor scheduling choices made by the random scheduler. Reliable SINR estimations should thus be achievable if candidate user groups are provided by a proper scheduling technique.

\subsection{PQS scheduler}
In the second evaluation step, the priority queue scheduler (PQS) presented in \cite{bib:STARDUST_D4.5} is implemented. In this context, we assign the $k$-th user a traffic request  $\mathcal{C}_k^{(REQ)}\in [\mathcal{C}_{min}, \mathcal{C}_{max}]$, $k=1,\dots,N_{UE}$, based on the population density in the user's area. The PQS algorithm selects groups of $N_{UE, sched}\leq N_B$ users to be served by 1) classifying the packets to be transmitted into $N_{PC}\geq 2$ priority classes based on the unmet capacity request and the remaining visibility of the corresponding user; 2) filling $N_{PC}$ priority-based queues with the corresponding packets; and 3) for each time slot in a scheduling period, randomly selecting $N_B$ packets from the non-empty queue with highest priority, ensuring that the selected users' CSI vectors' correlation (CSI-based scheduling) or the inter-user distance (geographic scheduling) are under a fixed threshold. By implementing the PQS algorithm, we consider the case in which a scheduler initially selects tentative groups of users to be served during the same time slot for which the SINR should be assessed (\textit{e.g.}, to determine the expected rate for each user, or to select the group that achieves the highest the sum-rate). The PQS scheduler parameters and the additional simulation parameters with respect to Table \ref{tab:SINRest_parameters} are reported in Table \ref{tab:SINRest_PQSparams}; the CSI and GEO scheduling constraint threshold was selected from the values reported in \cite{bib:STARDUST_D4.5} based on the minimum and maximum requested capacity $\mathcal{C}_{min}$ and $\mathcal{C}_{max}$. It should be noted that the models are assumed to be deployed as is, \textit{i.e.}, without retraining on user reports with PQS scheduling. For this reason, due to the different SINR statistics with respect to the training dataset, the obtained estimator is biased; such bias is estimated once and subtracted from the models’ output.
\begin{table}[t]
    \centering
    \begin{tabular}{|l|c|}
        \hline
        \textbf{Parameter} & \textbf{Value} \\
        \hline
        Scheduling slot duration & $T_{slot}^{(sched)}=10$ ms \\
        \hline
        Scheduling periodicity & $T_{sched}=2$ s \\
        \hline
        Maximum residual visibility & $\sigma_{vis}=50$ slots \\
        \hline
        Unmet capacity factor & $\sigma_{cap}=2$ \\
        \hline
        Number of priority classes & $N_{PC}=2$ \\
        \hline
        Minimum capacity request & $\mathcal{C}_{min}=\{5, 20\}$ Mbps \\
        \hline
        Maximum capacity request & $\mathcal{C}_{min}=\{100, 500\}$ Mbps \\
        \hline
        \end{tabular}
    \caption{Configuration parameters for PQS.}
    \label{tab:SINRest_PQSparams}
\end{table}
\begin{figure}[t]
    \centering
    \includegraphics[width=0.8\columnwidth]{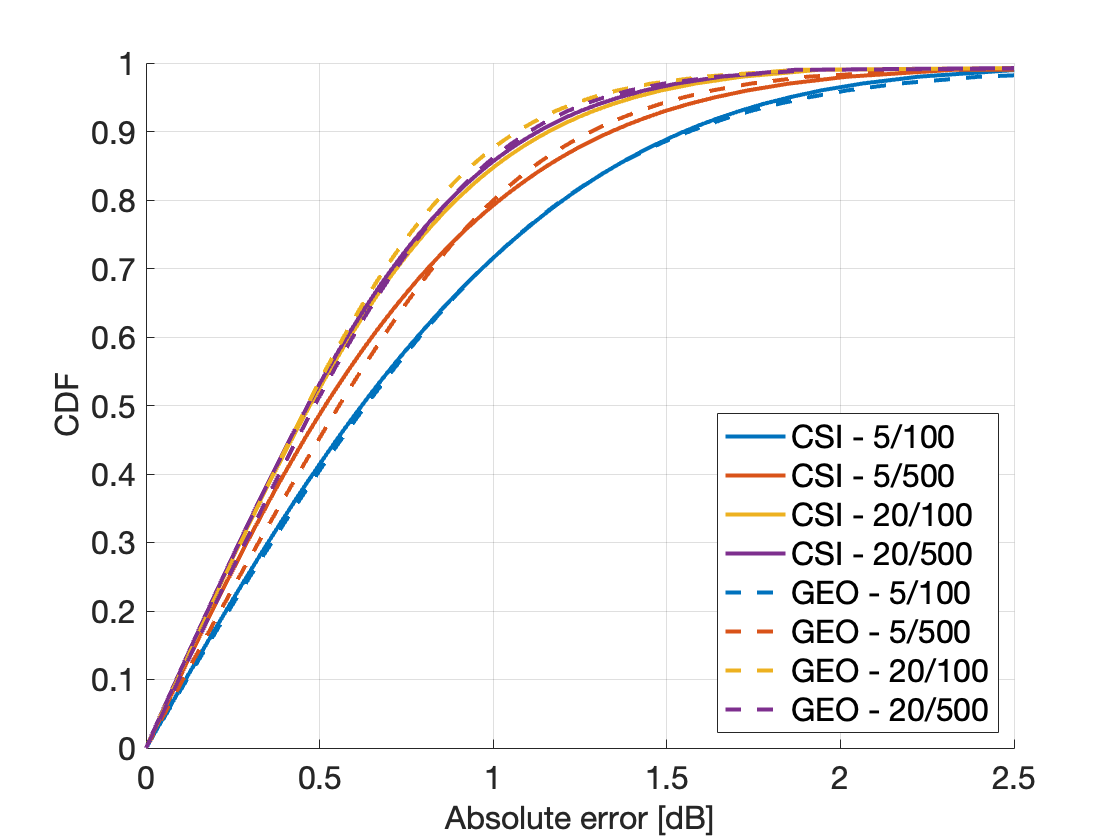}
    \caption{CDF of the SINR estimation absolute error with PQS scheduling for different capacity request ranges ($\mathcal{C}_{min}/\mathcal{C}_{max}$).}
    \label{fig:SINRest_CDFerror}
\end{figure}
Figure \ref{fig:SINRest_CDFerror} reports the CDFs of the absolute error $\mathbf{E}_{abs}=\left|\hat{\mathbf{Y}}^{(DMHSA)}-\mathbf{SINR}\right|$ achieved by the proposed models under the minimum and maximum capacity requests $\mathcal{C}_{min}$ and $\mathcal{C}_{max}$, respectively. The plot shows that both models are able to achieve a median absolute error under 0.7 dB for all cases. Overall, the performance of the DMHSA models is similar, with the location-based model providing slightly improved high percentiles and the CSI-based model excelling in the lower ones (except in the 5/100 case, where the opposite is true). The plot also shows that the cases with the largest $\mathcal{C}_{min}$ are the ones providing the best absolute error: as reported in \cite{bib:STARDUST_D4.5}, a low $\mathcal{C}_{min}$ value leads the PQS scheduler tends to schedule less users than the number of beams, thus resulting in consistently high SINR values, a case that was not well represented in the training data considering the implementation of a random scheduler; on the opposite, $\mathcal{C}_{min}=20$ Mbps results in a better aligned SINR distribution. Similarly, an increase of $\mathcal{C}_{max}$ from 100 Mbps to 500 Mbps with $\mathcal{C}_{min}=5$ Mbps leads to a significant improvement of the absolute error due to the more stringent constraint on the CSI vectors correlation and the inter-user distance. 

\begin{table}[t]
    \centering
    \begin{tabular}{|l|c|c|c|c|c|}
        \hline
         & & \multicolumn{2}{c|}{\textbf{CSI-DMHSA}} & \multicolumn{2}{c|}{\textbf{GEO-DMHSA}} \\
        \cline{3-6}
         & & \multicolumn{4}{c|}{$\mathcal{C}_{max}$ [Mbps]} \\
        \cline{3-6}
        & & \textbf{100} & \textbf{500} & \textbf{100} & \textbf{500} \\
        \hline
        \multirow{2}{*}{$\mathcal{C}_{min}$ [Mbps]} & \textbf{5} & 0.96 dB & 0.84 dB & 1.18 dB & 0.90 dB \\
        \cline{2-6}
         & \textbf{20} & 0.71 dB & 0.70 dB & 0.68 dB & 0.70 dB \\
        \hline
    \end{tabular}
    \caption{SINR estimation RMSE with PQS.}
    \label{tab:SINRest_RMSE}
\end{table}

Table \ref{tab:SINRest_RMSE} reports the SINR estimation RMSE achieved by the DMHSA models under the considered capacity requests. Notably, the RMSE remains under the 1dB threshold (except for the 5/100 case with GEO-DMHSA). The two models perform similarly at high $\mathcal{C}_{min}$; on the opposite, the CSI-DMHSA model has an advantage over its location-based counterpart at low $\mathcal{C}_{min}$. Clearly, the observations reported for Figure \ref{fig:SINRest_CDFerror} are also reflected in the RMSE.

\section{Conclusions}
In this paper, we proposed a DL model to estimate the SINR of a group of users scheduled for transmission in a user-centric beamforming NTN system, reducing the computational complexity with respect to the assessment of the SINR through the beamforming matrix while maintaining satisfying estimation accuracy.
The considered architecture, based on the MHSA mechanism, achieves Big-O complexity reduction of a factor 3 in the CSI-based case and of two orders of magnitude in the location-based case. This last result is a consequence of the lower dimensionality of the user reports for location-based user-centric beamforming systems, which avoid scaling the computational complexity by the number of antenna elements $N_R$. Furthermore, both models achieve satisfying SINR estimation performance considering PQS scheduling, with the RMSE not passing the 1dB threshold in most of the cases.
With the estimation performance being evaluated, the DMHSA models can be deployed for inference. Scheduling algorithms typically provide a single group of users to be served during the same time slot. While mechanisms, \textit{e.g.}, the priority queue in the PQS algorithm, are in place to improve the capacity offered in a time slot, the decision is often sub-optimal in this regard. Thus, the capacity to a group of potentially scheduled users may be evaluated through the SINR by means of DMHSA-based estimation. Future works will investigate SINR-estimation-aided schedulers, including the impact of the estimation accuracy on the top-K selection mechanism, as a potential solution to improve the achieved capacity in user-centric NTNs. Furthermore, online learning frameworks will be investigated to ensure the adaptability of the model to dynamic channel conditions.

\section{Acknowledgments}\label{Acknowledgment}
This work has been funded by the 5G-STARDUST project, which received funding from the Smart Networks and Services Joint Undertaking (SNS JU) under the European Union’s Horizon Europe research and innovation programme under Grant Agreement No 101096573. The views expressed are those of the authors and do not necessarily represent the project. The Commission is not liable for any use that may be made of any of the information contained therein.

\bibliographystyle{IEEEtran}
\bibliography{IEEEbib}

\end{document}